\documentclass[twocolumn,showpacs,preprintnumbers,amsmath,amssymb,prb]{revtex4}
\usepackage{epsfig}
\usepackage{graphicx}
\usepackage{dcolumn}
\usepackage{bm}

\begin{document}

\title{LDA+DMFT spectral functions and effective electron mass enhancement in superconductor LaFePO}

\author{S.~L.~Skornyakov,$^{1,2}$ N.~A.~Skorikov,$^{1}$ A.~V.~Lukoyanov,$^{1,2}$ A.~O.~Shorikov,$^{1,2}$ and V.~I.~Anisimov$^{1,2}$}
\affiliation{$^{1}$Institute of Metal Physics, Russian Academy of Sciences,
620990 Ekaterinburg, Russia \\
$^{2}$Ural State Technical University, 
620002 Ekaterinburg, Russia}

\date{\today}

\begin{abstract}
In this paper we report the first LDA+DMFT results (method 
combining Local Density Approximation with Dynamical Mean-Field 
Theory) for spectral properties of superconductor LaFePO. 
Calculated {\bf k}-resolved spectral functions reproduce recent 
angle-resolved photoemission spectroscopy (ARPES) data 
[D. H. Lu {\it et al}., Nature {\bf 455}, 81 (2008)]. Obtained 
effective electron mass enhancement values $m^{*}/m\approx$ 1.9 -- 2.2 
are in good agreement with infrared and optical studies 
[M.~M.~Qazilbash {\it et al}., Nature Phys. {\bf 5}, 647 (2009)], 
de Haas--van Alphen, electrical resistivity, and electronic specific 
heat measurements results, that unambiguously evidence for moderate 
correlations strength in LaFePO. Similar values of $m^{*}/m$ were 
found in the other Fe-based superconductors with substantially 
different superconducting transition temperatures. Thus, the 
dynamical correlation effects are essential in the Fe-based 
superconductors, but the strength of electronic correlations does 
not determine the value of superconducting transition temperature. 
\end{abstract}


\maketitle

\section{Introduction.} 
The discovery of superconductivity with the transition temperature 
$T_{c}\approx$~4~K in LaFePO\cite{Kamihara06} and $T_{c}\approx$ 
26--55 K in RO$_{1-x}$F$_x$FeAs (R = La, Sm)\cite{Kamihara08} has 
generated great interest to the new class of Fe-based 
superconductors.~\cite{review} While the microscopic mechanism of superconductivity 
in LaFePO is not yet clear,~\cite{Kuroki09,Yamashita09} its electronic 
properties have been studied extensively. Various experiments revealed 
the presence of electronic correlations in LaFePO. Analyzing angle-resolved 
photoemission spectroscopy (ARPES) data, D. H. Lu {\it et al}. (Ref.~\onlinecite{Lu}) 
demonstrated that the DFT band structure should be renormalized by a 
factor of 2.2 to fit the experimental angle-resolved photoemission 
spectra. From infrared and optical conductivity data the authors of 
Ref.~\onlinecite{Qazi} made a conclusion that the effective electron mass renormalization 
is about 2 in LaFePO. Similarly, the electron mass renormalization 
obtained from de Haas--van Alphen study is $m^{*}/m\approx$ 1.7 -- 2.1 (Ref.~\onlinecite{dHvA}) 
Comparison of the experimental electronic specific heat 
coefficient $\gamma_n$~=~10.1~mJ/mol~K$^2$ for LaFePO\cite{Suzuki09} 
with the DFT-value 5.9~mJ/mol K$^2$ (Ref.~\onlinecite{Lebegue}) gives the value of 
1.7 for electron mass enhancement. Also the electrical resistivity at 
low temperatures has $T^2$-dependence\cite{Suzuki09,Yamashita09} showing 
importance of correlation effects. However, the authors of Ref.~\onlinecite{Yang09} 
from analysis of x-ray absorption (XAS) and resonant inelastic x-ray 
scattering (RIXS) data for several iron pnictide compounds (SmO$_{0.85}$FeAs, 
BaFe$_2$As$_2$, LaFe$_2$P$_2$) arrived at a conclusion that correlations 
are not very strong here.

So far only a few electronic structure calculations for LaFePO by the 
DFT-based first-principles methods without any account for electronic 
correlations have been reported\cite{Lu,Lebegue,Vildosola,Che08,KamiPRB} 
thus making the problem of correlation effects study for this material 
very timely. 

The combination of Density Functional Theory (DFT) and Dynamical Mean-Field 
Theory (DMFT) called LDA+DMFT method\cite{LDA+DMFT} is presently 
recognized state-of-the-art many-particle method  to study correlation 
effects in real compounds. LDA+DMFT calculations for the FeAs-based 
superconductors\cite{Haule,jpcm,bafe,Aichhorn} lead to diverse conclusions 
on the strength of electron correlations in these materials. The 
authors of Ref.~\onlinecite{bafe} have proposed an extended classification 
scheme of the electronic correlation strength in the Fe-pnictides 
based on  analysis of several relevant quantities: a ratio of the 
Coulomb parameter $U$ and the band width $W$ ($U$/$W$), quasiparticle 
mass enhancement $m^{*}/m$, ${\bf k}$-resolved and ${\bf k}$-integrated 
spectral functions $A({\bf k},\omega)$ and $A(\omega)$. Applying this 
scheme to LDA+DMFT results for BaFe$_{2}$As$_{2}$ they came to conclusion 
that this material should be regarded as a {\it moderately} correlated metal. 
In this work we report the results of LDA+DMFT study for electronic 
correlation effects in LaFePO. For this purpose we have calculated 
spectral functions $A({\bf k}, \omega)$, effective electron mass enhancement 
$m^{*}/m$ and compare our results with the available measurements for 
LaFePO finding a very good agreement between calculated and experimental 
data. A moderate spectral functions renormalization corresponding to 
$m^{*}/m\approx$ 2 was found in LaFePO similar to values obtained for 
BaFe$_{2}$As$_{2}$\cite{bafe} while superconducting transition temperature 
in those materials could be an order of magnitude different.

\section{Method.} 
The LDA+DMFT scheme is constructed in the following way: First, a 
Hamiltonian $\hat H_{LDA}$ is produced using converged LDA results 
for the system under investigation, then the many-body Hamiltonian 
is set up, and finally the corresponding self-consistent DMFT equations 
are solved. By projecting onto Wannier functions,~\cite{projection} we 
obtain an effective 22-band Hamiltonian which incorporates five Fe {\it d}, 
three O {\it p}, and three P {\it p} orbitals per formula unit. In the 
present study we construct Wannier states for an energy window 
including both {\it p} and {\it d} bands. Thereby hybridization effects 
between {\it p} and {\it d} electrons were explicitly taken into account 
and eigenvalues of the Wannier functions Hamiltonian $\hat H_{LDA}$ exactly 
correspond to the 22 Fe, O, and P  bands from LDA.
The~LDA~calculations were performed with the experimentally determined 
crystal structure\cite{Zimmer} using the Elk full-potential linearized 
augmented plane-wave (FP-LAPW) code.~\cite{Elk} Parameters controlling 
the LAPW basis were kept to their default values. The calculated LDA band 
structure $\epsilon_{LDA}({\bf k})$ was found to be in good agreement 
with that of Leb\`egue \emph{et~al.} (Ref.~\onlinecite{Lebegue}).

The many-body Hamiltonian to be solved by DMFT has the form
\begin{equation}
\hat H= \hat H_{LDA}- \hat H_{dc}+\frac{1}{2}\sum_{i,\alpha,\beta,\sigma,\sigma^{\prime}}
U^{\sigma\sigma^{\prime}}_{\alpha\beta}\hat n^{d}_{i\alpha\sigma}\hat n^{d}_{i\beta\sigma^{\prime}},
\end{equation}
where $U^{\sigma\sigma^{\prime}}_{\alpha\beta}$ is the Coulomb interaction 
matrix, $\hat n^d_{i\alpha\sigma}$ is the occupation number operator for 
the $d$ electron with orbital $\alpha$ or $\beta$ and spin indices $\sigma$ 
or $\sigma^{\prime}$ in the $i$-th site. The term $\hat H_{dc}$ stands for 
the {\it d}-{\it d} interaction already accounted in LDA, so called 
double-counting correction. The double-counting has the form 
$\hat H_{dc}=\bar{U}(n_{\rm dmft}-\frac{1}{2})\hat{I}$ where $n_{\rm dmft}$ 
is the total self-consistent number of {\it d} electrons obtained within 
the LDA+DMFT and $\bar{U}$ is the average Coulomb parameter for the {\it d} shell. 

The DMFT self-consistency equations were solved iteratively for 
imaginary Matsubara frequencies. The auxiliary impurity problem 
was solved by the hybridization function expansion Continuous-Time Quantum 
Monte-Carlo (CTQMC) method.~\cite{QMC} In the present implementation 
of the CTQMC impurity solver the Coulomb interaction is taken into account 
in density-density form. The elements of $U_{\alpha\beta}^{\sigma\sigma'}$ 
matrix were parameterized by $U$ and $J$ according to procedure 
described in Ref.~\onlinecite{LichtAnisZaanen}. We used interaction parameters 
$\bar{U}$~=~3.1~eV and $J$~=~1~eV similar to the values calculated 
by the constrained LDA method for Wannier functions\cite{Korotin} 
in Fe-pnictides.~\cite{jpcm} Calculations were performed in the 
paramagnetic state at the inverse temperature $\beta=1/T$ = 20 eV$^{-1}$. 
The real-axis self-energy needed to calculate spectral 
functions was obtained by the Pad\'e approximant\cite{Pade} (see Appendix).

\section{Results and discussion.} 
\begin{figure}[h]
\centering \vspace{0.0mm}
\includegraphics[width=0.95\linewidth]{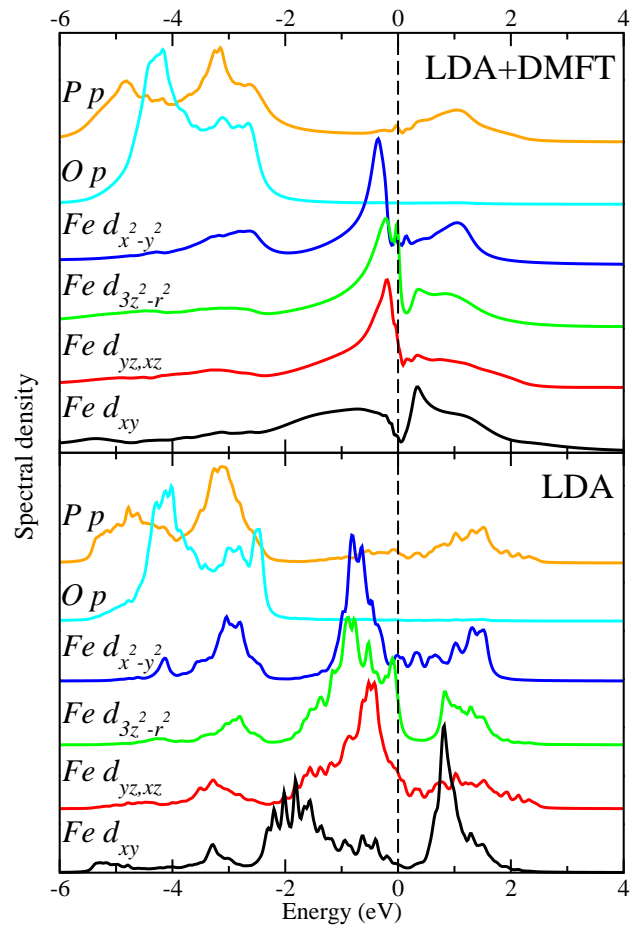}
\caption{(Color online) Orbitally resolved Fe~3{\it d}, O~{\it p} and P~{\it p} 
normalized spectral functions of LaFePO obtained within LDA+DMFT (upper panel) 
are compared with the LDA results (lower panel).}
\label{DMFTvsLDA}
\end{figure}
The orbitally resolved Fe~3$d$, O~{\it p} and P~{\it p} spectral
functions computed within LDA and LDA+DMFT, respectively, are compared in 
Fig.~\ref{DMFTvsLDA}. Within the LDA all five Fe $d$ orbitals form a common 
band in the energy range ($-$2.5, $+$2.0)~eV relative to the Fermi level 
(band width $W\approx$  4.5~eV). There is a significant hybridization of 
the Fe $3d$ orbitals with the P~$p$ and O~$p$ orbitals, leading to appearance 
of Fe $d$ states contribution in the energy interval ($-$5.5, $-$2.5)~eV where 
the P $p$ band is located. The corresponding features of LDA+DMFT spectral 
functions (upper panel in Fig.~\ref{DMFTvsLDA}) in the energy area 
($-$5.5, $-$2.5)~eV should not be mistaken for Hubbard bands because the 
same peaks are present in non-correlated LDA bands (lower panel in 
Fig.~\ref{DMFTvsLDA}). Correlation effects do not result in Hubbard bands 
appearance but lead to significant renormalization of the spectral function 
around the Fermi energy: ``compressing'' of energy scale so that separation 
between peaks of LDA+DMFT curves becomes $\approx$2 times smaller than in 
corresponding non-correlated spectra.

\begin{figure}[h]
\centering \vspace{0.0mm}
\includegraphics[width=0.885\linewidth]{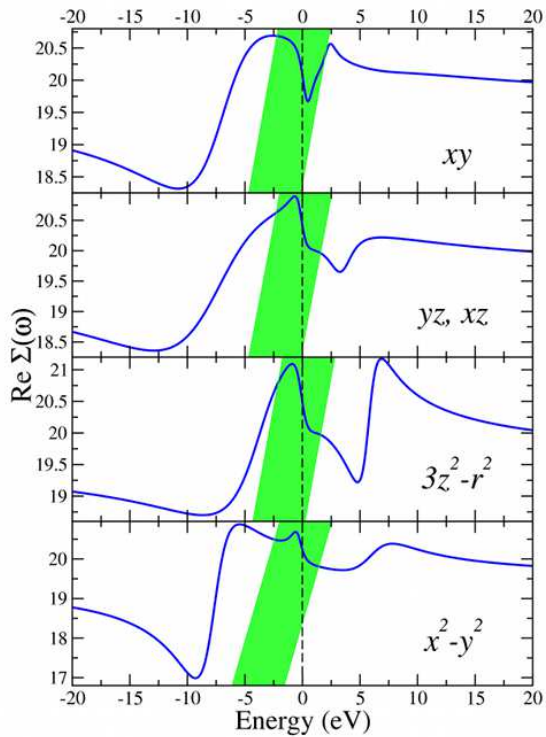}
\caption{(Color online) The real part of the self-energy $\Sigma(\omega)$ (black line) depicted 
together with $\omega+(H_{dc})_{ii}$ (light stripe, see text).}
\label{Sigmax2my2}
\end{figure}

It is instructive to plot energy dependence of real part of self-energy 
Re$\Sigma(\omega)$ (see Fig.~\ref{Sigmax2my2}). Peaks in spectral function 
$A({\bf k}, \omega)$ are determined by the poles of 
$(\omega-\epsilon({\bf k})-\Sigma(\omega))^{-1}$ function or the energy values
$\omega=\epsilon({\bf k})+\mbox{Re}\Sigma(\omega)$ (here $\epsilon({\bf k})$ 
is non-correlated band dispersion). In Fig.~\ref{Sigmax2my2} together with 
Re$\Sigma(\omega)$ a function $\omega+(H_{dc})_{ii}$ is  plotted as a stripe 
having the width of non-correlated band $\epsilon({\bf k})$. 
The peaks of spectral function $A({\bf k}, \omega)$ correspond to energy area 
where this stripe crosses Re$\Sigma(\omega)$ curve. As one can see such 
crossing happens only once in the energy interval around the Fermi level so 
that spectral functions will have only poles corresponding to quasiparticle 
bands and no Hubbard band poles will be observed.

A quantitative measure of the electron correlation strength is 
provided by the quasiparticle renormalization factor
$Z=(1-\frac{\partial \Sigma}{\partial\omega}|_{\omega=0})^{-1}$
which gives an effective mass enhancement $m^{*}/m=Z^{-1}$. In
general, the self-energy is a matrix, leading to different effective
masses for different bands. The calculated $m^{*}/m$ values for 
every {\it d}-orbital are presented in Table 1. The $d_{x^{2}-y^{2}}$ 
orbital has the smallest effective mass renormalization $m^{*}/m$ = 1.942. 
The other {\it d} orbitals have approximately the same value $m^{*}/m\approx$~2.2. 
\begin{table}[h]
\caption{Effective mass renormalization $m^{*}/m$ of quasiparticles in
LaFePO for different orbitals of the Fe {\it d} shell 
from the LDA+DMFT calculation.}
\vspace{3.0mm} \centering
\begin{tabular}{c|cccc}
\hline \hline
Orbitals  & $d_{xy}$ & $d_{yz,xz}$ & $d_{3z^{2}-r^{2}}$ & $d_{x^{2}-y^{2}}$ \\
\hline
$m^{*}/m$  & 2.189   &    2.152    & 2.193              & 1.942              \\
\hline \hline
\end{tabular}
\end{table}

The calculated effective mass enhancement $m^{*}/m\approx$~1.9 -- 2.2 in 
LaFePO agrees very well with the de Haas--van Alphen experiments\cite{dHvA} 
where it was found to range from 1.7 to 2.1, and with the estimations of 
the effective mass renormalization of a factor of 2 from optical conductivity 
data\cite{Qazi} and specific heat measurements.~\cite{Suzuki09} 

\begin{figure}[h]
\centering \vspace{0.0mm}
\includegraphics[width=0.95\linewidth]{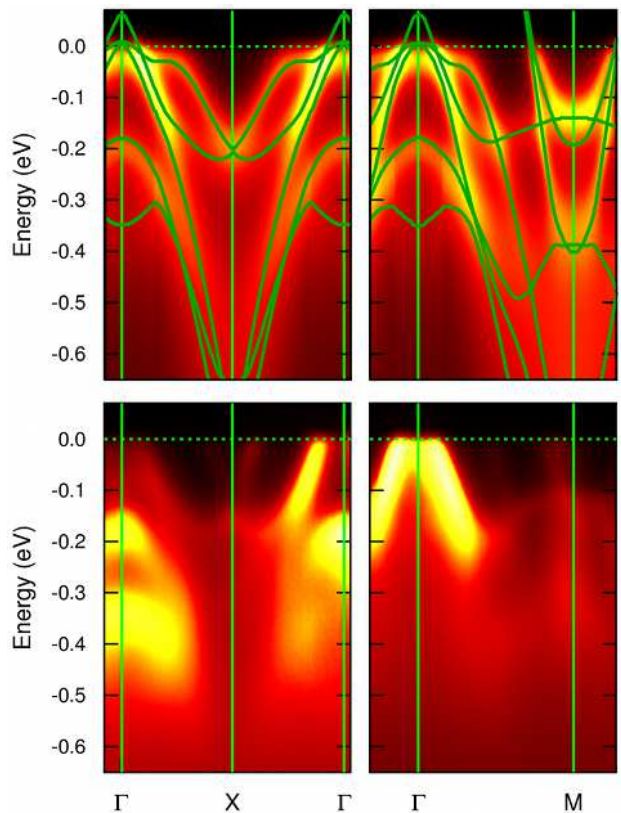}
\caption{(Color online) The {\bf k}-resolved total spectral function
A({\bf k}, $\omega$) of LaFePO along the $\Gamma-X-\Gamma$ and $\Gamma-M$ 
lines in the Brillouin zone is depicted as a contour plot. Upper panel: The LDA+DMFT 
spectral function. Lower panel: The corresponding experimental ARPES 
intensity map of Lu~{\it et~al}.~(Ref.~\onlinecite{Lu}).} 
\label{Ake_contour}
\end{figure}

We now calculate the {\bf k}-resolved spectral function
\begin{equation}
A({\bf k}, \omega)=-{\rm Im}\frac{1}{\pi} Tr[(\omega+\mu)\hat{I}-\hat h_{\bf k}-\hat{\Sigma} (\omega)]^{-1}.
\label{SFunction}
\end{equation}
Here $\hat h_{\bf k} = \hat H_{LDA}- \hat H_{dc}$ is the 22$\times$22 Hamiltonian 
matrix on a mesh of {\bf k}-points and $\mu$ is the self-consistently determined
chemical potential. In Fig.~\ref{Ake_contour} we compare our results with ARPES 
data of Lu {\it et al}. (Ref.~\onlinecite{Lu}). Both theory and experiment show dispersive 
bands crossing the Fermi level near the $\Gamma$ and M points. In addition, two 
bands can be seen at --0.2 eV and in the region from --0.3 to --0.4 eV near the 
$\Gamma$ point. The calculated shape and size of the hole and electron pockets 
centered at the $\Gamma$ and M points, respectively, are in good agreement with 
the ARPES, see Fig.~\ref{Ake_contour} (lower panel), and de~Haas--van~Alphen\cite{dHvA} data.

The correlated band structure $\epsilon_{DMFT}({\bf k})$ 
is also shown in Fig.~\ref{Ake_contour} (upper panel). Near the Fermi energy, 
i.e., in the energy range from --0.2 eV to zero where quasiparticles 
are well defined (as expressed by a linear behavior of Re$ \Sigma(\omega)$, 
see Fig.~\ref{Sigmax2my2}), this dispersion is very well represented 
by the scaling relation 
$\epsilon_{DMFT}({\bf k})=\epsilon_{LDA}({\bf k})/(m^*/m)$, 
with $m^*/m$ taken as the computed mass enhancement from Table~1.

The present results mean that the band structure in LaFePO is renormalized 
by the correlations and consists of the quasi-two-dimensional Fermi surface 
sheets observed in the experiments. At the same time,  there is no substantial 
spectral weight transfer from quasiparticle bands near the Fermi energy to 
Hubbard bands and the system is far away from metal-insulator transition. Such 
spectral function behavior does not allow to classify LaFePO either as {\it strongly} 
or {\it weakly} correlated material. According to classification proposed in 
Ref.~\onlinecite{bafe} this material can be regarded as a {\it moderately} 
correlated metal, like BaFe$_{2}$As$_{2}$.~\cite{bafe}

In other iron pnictide materials the electron mass enhancement 
was also reported to be close to value of $\approx$2. For example, from the 
de Haas--van Alphen experiments $m^{*}/m$ = 1.13 -- 3.41 (Ref.~\onlinecite{Analytis}) 
was found for SrFe$_{2}$P$_{2}$, isostructural analogue of the superconducting 
compounds Sr$_{1-x}$K$_{x}$Fe$_{2}$As$_{2}$ ($T_{c}$ = 37 K). For the series 
of isostructural compounds Ba(Fe$_{1-x}$Co$_{x}$)$_2$As$_{2}$, $x$ = 0 -- 0.3  
the authors of Ref.~\onlinecite{Brouet} evaluated $m^{*}/m$ = 2 -- 4 from the ARPES data. 
Also from the analysis of ARPES an electron mass renormalization by 2.7 was evaluated 
in Ba$_{0.6}$K$_{0.4}$Fe$_{2}$As$_{2}$ ($T_{c}$ = 37 K)\cite{YiLu} as comparing 
with the DFT-band structure. Further, the DFT-calculated plasma frequencies 
are by a factor of 1.5 to 2 larger than the experimental values,~\cite{Drechsler} 
showing the electron mass enhancement of the same strength for LaFePO, LaFeAsO, 
SrFe$_{2}$As$_{2}$, BaFe$_{2}$As$_{2}$, 
K$_{0.45}$Ba$_{0.55}$Fe$_{2}$As$_{2}$ ($T_{c}\approx$~30~K), 
and LaO$_{0.9}$F$_{0.1}$FeAs ($T_{c}\approx$ 26 K). From the reported data, it 
follows that the electronic correlation strength in the FeAs-based superconductors 
with the superconducting transition temperatures up to 37 K and their parent 
compounds is the same as in superconductor LaFePO with $T_{c}\approx$~4~K. 
Thus, the strength of electronic correlations in the Fe-based superconductors seems 
to be not intrinsically connected with the superconducting transition temperature.

\section{Conclusion.} 
By employing the LDA+DMFT method we have calculated the spectral functions 
and single-particle {\bf k}-resolved spectrum of superconductor LaFePO for the 
first time. Very good agreement with the ARPES data was found. In the spectral 
functions we observed no substantial spectral weight transfer. The obtained 
effective electron mass enhancement values $m^{*}/m\approx$ 1.9 -- 2.2 are in 
good agreement with infrared and optical studies, de Haas--van Alphen, and 
specific heat results. The electronic correlation strength in LaFePO with small 
value of superconducting temperature 4 K is similar to the other Fe-pnictide 
superconductors with the transition temperatures up to 37 K.

\section{Acknowledgments.} 
The authors thank D. Vollhardt for useful discussions, J.~Kune\v{s} for providing 
DMFT computer code used in our calculations, P.~Werner for the CT-QMC impurity 
solver, D.~H.~Lu and Z.-X.~Shen for their ARPES data. This work was supported 
by the Russian Foundation for Basic Research (Projects Nos. 10-02-00046a, 09-02-00431a, 
and 10-02-00546a), the Dynasty Foundation, the fund of the President of the Russian 
Federation for the support of scientific schools NSH 4711.2010.2, the Program of
the Russian Academy of Science Presidium ``Quantum microphysics of condensed 
matter'' N7, Russian Federal Agency for Science and Innovations (Program 
``Scientific and Scientific-Pedagogical Trained of the Innovating Russia'' for 
2009-2010 years), grant No. 02.740.11.0217. S.L.S. and V.I.A. 
are grateful to the Center for Electronic Correlations and Magnetism, University 
of Augsburg, Germany for the hospitality and support of the Deutsche 
Forschungsgemeinschaft through SFB 484.

\section{Appendix: self-energy on the real axis}

\begin{figure}[!b]
\centering \vspace{0.0mm}
\includegraphics[width=0.88\linewidth]{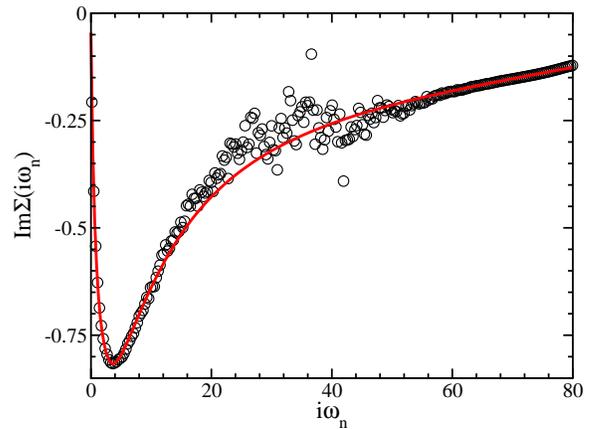}
\caption{(Color online) The imaginary part of the Fe $d_{yz}$ self-energy $\Sigma(i\omega_{n})$ 
calculated within LDA+DMFT (black circles) is compared with the corresponding Pad\'e approximant
(red curve).} 
\label{DMFTvsPade}
\end{figure}

In the DMFT method temperature (Matsubara) Green functions formalism is used with arguments 
in the form of imaginary time $\tau$ or corresponding Matsubara imaginary energies 
$i\omega_n=i(2n+1)\pi/\beta$. In order to calculate a spectral function one needs to have 
self-energy as a function of real energy $\Sigma(\omega)$ that means to perform analytical 
continuation of the function $\Sigma(i\omega_n)$ to real axis. One of the usual algorithms 
for analytical continuation is Pad\'e approximant,~\cite{Pade} and it was successfully used 
in earlier LDA+DMFT calculations where effective impurity problem was solved by Iterative 
Perturbation Theory (IPT) method.~\cite{LDA+DMFT_JPCM} However, when impurity problem is solved 
by stochastic Quantum Monte Carlo (QMC) method numerical noise appears in calculated 
$\Sigma(i\omega_n)$, see Fig.~\ref{DMFTvsPade}. Attempts to apply the Pad\'e approximant 
method to such noisy data result in completely wrong widely oscillating real axis function. 

In order to solve this problem Maximum Entropy method (MEM) method\cite{jarrell96} was proposed. 
In this method spectral function $A(\omega)$ corresponding to imaginary time Green function 
$G(\tau)$ from QMC calculation is found as a best approximation to solution of the integral equation:
\begin{equation}
G(\tau)=-\int_{-\infty}^{\infty}d\omega\frac{e^{-\tau\omega}}{1+e^{-\beta\omega}}A(\omega)
\ ,
\label{ME8}
\end{equation}
with the condition of maximization of effective entropy functional that gives a smooth spectral 
function. The resulting spectral function $A(\omega)$ is identified then with the {\bf k}-integrated 
analogue of Eq.~\ref{SFunction} that gives equations for unknown self-energy $\Sigma(\omega)$. 
For many-orbital case that gives a set of equations with the corresponding number of unknown 
variables $\Sigma_{i}(\omega)$. Solution of such a set of equations can be a rather difficult 
problem. In addition to that the MEM method smears out all high-energy features in $A(\omega)$ 
due to the factor $e^{-\beta\omega}$ in the kernel of integral equation~\eqref{ME8}.

\begin{figure}[!t]
\centering \vspace{0.0mm}
\includegraphics[width=0.8\linewidth]{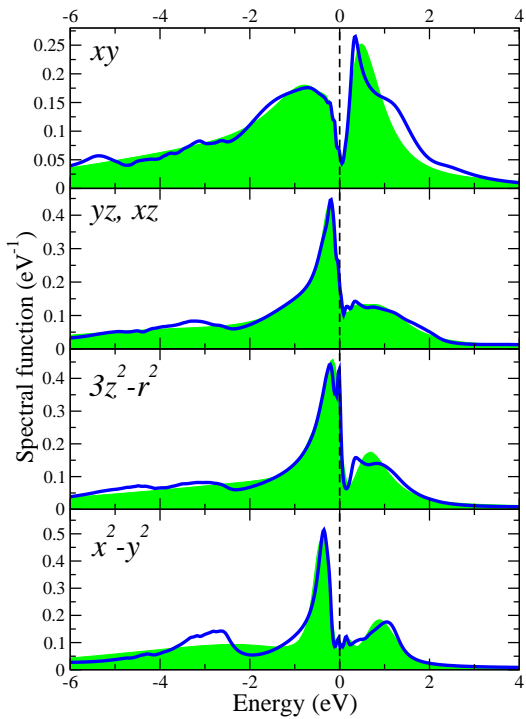}
\caption{(Color online) Orbitally resolved Fe~3{\it d} spectral functions of LaFePO from 
the Maximum Entropy method (green shaded areas) and the real-axis self-energy $\Sigma(\omega)$ 
obtained with the use of Pad\'e approximation (blue curves).} 
\label{MEvsPade}
\end{figure}

In the present work we have used a modified version of the Pad\'e approximant method.
To make the analytical continuation procedure of the noisy self-energy $\Sigma(i\omega)$ 
numerically stable, we construct the approximant using only those  frequencies values where 
the self-energy is a smooth function. Practically, that means to use a few first $i\omega_n$ 
at the lowest frequencies and  the data at the large frequencies where  $\Sigma(i\omega)$ 
approaches asymptotic behavior. In the result we obtain a smooth function that has correct 
analytical behavior at two limits: small energies $\omega\rightarrow 0$ and large energies 
where it obeys known asymptotic.  

In Fig.~\ref{DMFTvsPade} we compare the Fe $d_{yz}$ self-energy $\Sigma(i\omega_{n})$ obtained 
within Pad\'e approximation with the corresponding numerical data from QMC solution of DMFT 
equations. The approximant has accurate derivatives in vicinity of zeroth Matsubara frequency, 
that guaranties correct analytical properties near the Fermi level and correct asymptotic behavior. 
At the same time it approximates the noisy region with a smooth curve. 

In Fig.~\ref{MEvsPade} the Fe 3{\it d} spectral functions obtained with Pad\'e real-axis 
self-energy are compared with the MEM curves. The results for energies near the Fermi level 
are in very good agreement with each other. However, going to the higher and lower energies MEM 
curve very soon becomes smeared and nearly featureless while the curve obtained with Pad\'e real-axis 
self-energy has much better resolved peaks and shoulders. For example, all the peaks in the energy 
region ($-$6, $-$2)~eV are completely missed in  MEM curve. This is due to exponential
nature of the MEM kernel that suppresses all high-energy features.

\end{document}